\documentclass[10pt]{emulateapj}
\newcommand{\beq}{\begin{equation}}
\newcommand{\eeq}{\end{equation}}
\newcommand{\bea}{\begin{array}}
\newcommand{\eea}{\end{array}}

\shorttitle{Configuration of KOI-152 System} \shortauthors{Su Wang et al.}

\begin{document}

\title{Predicting the Configuration of Planetary System: KOI-152 Observed by Kepler}

\author{Su Wang\altaffilmark{1}, Jianghui Ji\altaffilmark{1}, and Ji-lin Zhou\altaffilmark{2}}

\altaffiltext{1}{Purple Mountain Observatory, Chinese Academy of
Sciences, Nanjing 210008, China; wangsu@pmo.ac.cn, jijh@pmo.ac.cn.}

\altaffiltext{2}{Department of Astronomy \& Key Laboratory of Modern
Astronomy and Astrophysics in Ministry of Education, Nanjing
University, Nanjing 210093, China.}

\begin{abstract}
The recent Kepler discovery of KOI-152 reveals a system of three hot
super-Earth candidates that are in or near a 4:2:1 mean motion
resonance.  It is unlikely that they formed in situ, the planets
probably underwent orbital migration during the formation and
evolution process. The small semimajor axes of the three planets
suggest that migration stopped at the inner edge of the primordial
gas disk. In this paper we focus on the influence of migration
halting mechanisms, including migration "dead zones", and inner
truncation by the stellar magnetic field. We show that the stellar
accretion rate, stellar magnetic field and the speed of migration in
the proto-planetary disk are the main factors affecting the final
configuration of KOI-152. Our simulations suggest that three planets
may be around a star with low star accretion rate or with high
magnetic field. On the other hand, slow type I migration, which
decreases to one tenth of the linear analysis results, favors
forming the configuration of KOI-152. Under such formation scenario,
the planets in the system are not massive enough to open gaps in the
gas disk. The upper limit of the planetary masses are estimated to
be about $15,~19$, and $24~M_\oplus$, respectively. Our results are
also indicative of the near Laplacian configurations that are quite
common in planetary systems.
\end{abstract}


\section{Introduction}
The Kepler Mission, launched in March 2009, photometrically monitors
a large patch of sky with sufficient precision to detect terrestrial
sized planets in potentially habitable orbits.  The Kepler Mission
has released their first 16 months data of 2321 transiting planet
candidates \citep{Bata12,Fab12}. The mission is sensitive to a
larger range of semimajor axes than ground-based transit surveys
\citep{Borucki10}. Thus there are opportunities to detect
multiplanetary systems. According to statistical results on the
first four months data, $\sim$ 17\% systems have multiple planet
candidates \citep {Borucki11}. More than a dozen multiple transiting
planet systems will be confirmed or rejected by means of transit
timing variations (TTVs) \citep{Ford11}.  The candidate multiple
planetary systems show that at least $\sim$ 16\% contain a pair of
planets close to 2:1 period commensurability (with a period ratio of
two planets ranging from 1.83 to 2.18) \citep{Lissauer11}.
Furthermore, the detailed statistical analysis of the release data
over 16 months suggests that the fraction of multiple planetary
systems has raised from 17\% to 20\% and the number of the resonant
systems also increases \citep{Bata12}. The resonant systems are of
value to the researchers, who investigate formation and evolution of
the planetary systems. \citet{steffen10} analyzed five Kepler target
stars and their planet candidates, in which KOI-152 (KOI: Kepler
Objects of Interest) is a known system, consisting of three
planetary candidates. Moreover, three planets, orbiting about an F
dwarf star (KIC 8394721) with a mass of $1.4~ M_\odot$, are very
close to 4:2:1 mean motion resonance (MMR) \citep{steffen10}. Table
1 shows the orbital elements of the planets in some detail, i.e.,
152.01, 152.02, and 152.03 in the order that the transit detection
software identified them in the Kepler data, and hereafter we label
them as Planet 01, 02, and 03, respectively, for brevity. Their
masses are now estimated to be in the range of (20-100) $M_\oplus$,
(9-30) $M_\oplus$, and (9-30) $M_\oplus$, respectively. Considering
the density limit and formation scenarios, observers may assume them
to be 60 $M_\oplus$, 15 $M_\oplus$, and 15 $M_\oplus$, respectively.
Although the semimajor axis of Planet 01 is not well determined,
estimated value indicates that they are very close to the central
star. The orbital period ratios of each pairs are
$P_{02}/P_{03}\simeq 2.033$ and $P_{01}/P_{02}\geq 1.896$. The
eccentricities are estimated to be zero and no evidence for large
eccentricities has been revealed \citep{steffen10}. Therefore, we
assume the eccentricities to be zero in this work. Based on the
possibility of observing such a three-planet system, the planets
likely occupy nearly coplanar orbits with a small deviation of
inclination from fundamental framework \citep{steffen10}.

It is well-known that the planetary configurations involved in 4:2:1
MMR, also occur both in our solar system and exoplanetary systems.
For instance, the Galilean moons of Jupiter had been revealed in a
three-body Laplace resonance over several hundred years. In
addition, another example is that three super-Earths constitute the
HD 40307 system, where three planets are near 4:2:1 MMRs
\citep{PT10}, similar to KOI-152. For HD 40307, three proto-planets
formed in the proto-planetary disk with configuration near MMRs,
which strongly constrains the planetary formation and orbital
migration theories. In this sense, the formation scenario of such
near Laplacian configuration is of great interest to the
researchers, and such investigations may test some planetary
formation theory.

According to core-accretion model, a planet with semimajor axis $a$
can grow up with the material surrounding it to mass $m_{\rm iso}$
\citep{idalin04}
\begin{equation}
M_{\rm{iso}}\simeq 0.16f_{\rm d}^{3/2}\gamma_{\rm{ice}}^{3/2}
\left(\frac{\triangle_{\rm{fz}}}{10R_{\rm
h}}\right)^{3/2}\left(\frac{a}{\rm{AU}}\right)^{3/4}\left(\frac{M_*}{M_\odot}\right)^{-1/2}M_\oplus,
\end{equation}
where $R_{\rm h}=(m/3M_*)^{1/3}a$ is the Hill radius of a planet
with a mass $m$, and $\triangle_{\rm fz}\sim 7-10$ $R_{\rm h}$ is
the so-called feeding zone of the planet, $M_*$ is the mass of
central star, $f_{\rm d}$ is the enhancement factor to the Minimum
Mass Solar Nebula (MMSN), $\gamma_{\rm ice}$ is the volatile
enhancement for exterior or interior to the snow line $a_{\rm ice}$
with a value of 4.2 or 1, respectively. For KOI-152, we adopt
$a_{\rm ice}=5.29$ AU \citep{idalin04}. If three planets formed in
the region as shown in Table \ref{tbl-1} with their estimated
minimum masses, $f_{\rm d}$ should  be at least 37. Hence, we may
conclude that it is impossible that all planets formed \textit{in
situ}.

Now, it is widely believed that there are mainly two formation
mechanisms to produce short-period planets, e.g., planet-planet
scattering and planetary migration \citep{RF96}. The planet-planet
scattering scenario always requires a massive planet to stir up the
eccentricities of other bodies and trigger the scattering process.
After this process had done, some planets may obtain high
eccentricity \citep{RF96}, away from their initial locations.
However, the planets of KOI-152 are near 4:2:1 MMR with nearly
circular orbits, and in this sense it seems to be impossible to tune
scattering scenario to yield such a precise configuration.

Another mechanism is that the orbital migration occurs in the
gaseous disk \citep{GT80,lin96}. Multiple planets are likely to be
captured into MMRs given an appropriate migration speed
\citep{MS01}. As MMR commonly emerges in the planetary systems, the
orbital migration is now considered as one of the plausible
mechanisms to form such systems. Take the GJ 876 system as an
example, consisting of four planets, GJ 876 b and GJ 876 c are
locked into 2:1 MMR, which can be explained by migration scenario
\citep{lee02,Ji02,Ji03,Zhou05,Zhang10}.

Recently, an alternative scenario of collision-merger has been
proposed to account for the formation of short-period planets
\citep{Ji11}. In this mechanism, several embryos can be excited by
giant planets after the gas of the disk depletes, and then merge
into a larger body moving on a close-in orbit. Nevertheless, in
these scenarios, planets are not easy to form in MMRs.

Considering the aforementioned factors, we propose a scenario to
produce a configuration of KOI-152. First, three planets are assumed
to have formed in the region away from the star with their nominal
masses. Then, the planets may undergo type I or II migration due to
interaction with the gaseous disk until they halt at the inner
region of the disk. In this phase, three planets are trapped into
MMRs during the migration. Finally, tidal effects, arising from the
central star, circularizes their orbits.

Such a scenario had been supposed by \citet{Ter07}. They calculated
a series of runs that are composed of 10-25 planets or planetary
cores in a disk with masses ranging from 0.1 to 1 $M_\oplus$, which
undergo type I migration. They showed that hot super-Earths or
Neptunes do not become isolated during their inward migration, and
the companions on near-commensurate orbits always survive.
Nevertheless, in their work, the authors did not consider the effect
of different speed of type I migration, which can occur. In a more
recent work, \citet{PT10} investigated the formation of the HD 40307
system, bearing a resemblance to KOI-152. They adopted a similar
formation scenario and found that in the end of the simulations, the
planets are driven out of Laplace resonances due to tidal effects
with the central star, and they finally reach a planetary
configuration very close to HD 40307. Furthermore, \citet{WZ11}
accounted for the contribution of type I migration and mainly
focused on the speed of migration under the perturbation of
gas-giants in the outer region in M dwarf system. They further point
out that pairs of planets favor forming not only near the 2:1 MMR
but other first order resonances.

One of the important factors that influence final configurations is
holes in the gas disk. In general, holes can appear at two
positions. One is the boundary of the dead zone and the active zone
in the midplane of the disk responding to the magnetorotational
instability (MRI). For protostellar gas disk in ad hoc
$\alpha$-prescription \citep{SS73}, the mass accretion rate, $\dot
{M_g}=3\pi\alpha c_s h \Sigma_g$, is constant across the whole disk,
where $h=c_s/\Omega_k$ is the disk scale height, $c_s$, $\Omega_k$,
and $\Sigma_g$ are the sound speed of the midplane, the Kepler
angular velocity, and the gas density, respectively. As the value of
$\alpha$ decreases from the active zone ($\sim$ 0.018; Sato et al.
2000) to the dead zone ($\sim$ 0.006), in the midplane the value of
the gas density in the dead zone (outer) is three times of that in
the active zone (inner). Then a maximum density location occurs.
Another is the inner hole of the disk caused by the coupling of the
star's magnetic field with the gas. Gas falls onto the surface of
the central star under the effect of the torque induced by the
stellar magnetic field, a truncation happens at the inner region and
an inner hole appears. A maximum density of the gas disk appears at
the boundary of the inner hole. Due to the variation of density, the
speed of type I migration may be changed. The two mentioned regions
play a significant role in forming the final planetary
configuration, especially for low-mass planetary systems.

In this work, we focus on exploring the configuration formation of
KOI-152, mainly on the following aspects: (a) the speed of type I
migration of planets, (b) the density profile of gas disk, (c) the
possible range of masses of three planets, and (d) the nature of the
star in the system. This paper is organized as follows: In \S 2, we
introduce the adopted disk model, orbital migration and eccentricity
damping models in the investigation.  In \S 3., we present the
simulation method and outcomes. Finally, we conclude and summarize
the results in \S 4.

\begin{table*}
\centering \caption{Orbital parameters of the KOI-152 planetary
system. \vspace{0.5cm} \label{tbl-1}}
\begin{tabular*}{10cm}{@{\extracolsep{\fill}}ccccc}
\tableline
  ID & Semimajor axis  & Period &   Eccentricity   & Mass\\
& (AU) & (day) && $(M_{\oplus})$\\
\tableline
152.03& 0.124 & 13.48 &   0.00 &  15 \\
152.02& 0.199 & 27.40 &   0.00 &  15 \\
152.01& 0.305 & $\geq$ 51.94 &   0.00 &  60\\
\tableline
\end{tabular*}

\end{table*}

\section{Models}
\subsection{Disk Model}
In order to explore the configuration formation, we consider a
system consisting of a central star and three planets, which formed
far away from the star. We assume that three planets are initially
embedded in a gaseous disk. The surface density is given as
\citep{pringle81}
\begin{equation}
\Sigma_{\rm g}=\frac{\dot {M}}{3\pi\nu(a)},
\end{equation}
where $\dot {M}$ is the accretion rate of the star and $\nu (a)$ is
the effective viscosity at the orbit of a semimajor axis \textit{a}.
According to the observation of young cluster $\rho$-Oph, the
accretion rate of the star can be written as \citep{natta,Vor09}
\begin {equation}
\dot M\simeq 2.5\times
10^{-8}\left(\frac{M_*}{M_\odot}\right)^{1.3\pm 0.3}M_\odot~ {\rm yr}^{-1}.
\label{mdot}
\end {equation}

\begin{figure}
\begin{center}
  \epsscale{1.20}\plotone{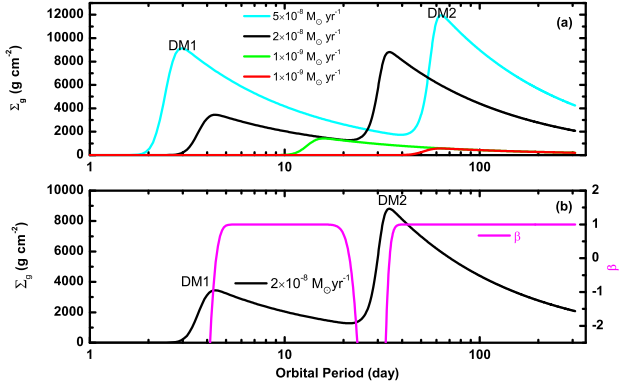}
 \caption{ The density profile of the gas disk and the values of $\beta$. The upper panel (a) shows
 the gas density profile with star accretion rate ($\dot M$ = $[1\times 10^{-9}, 5\times 10^{-8}]~M_\odot ~{\rm yr^{-1}}$), $\alpha_{\rm dead}$ = 0.001, and $\alpha_{\rm
 mri}$ = 0.01. The parameters of the star are set to be $B_*$ = 0.5 KG and
 $R_*$ = 2.5 $R_\odot$ except the red line for the cases in Group 4 with high magnetic field $B_*$ =2.5 KG. The bottom panel (b) illustrates the values of $\beta$, taking
 $\dot M$ = $2\times 10^{-8}~M_\odot~ {\rm yr^{-1}}$ as an example.
  Two maximum density locates at $\sim$ 3.6 days and 26 days, respectively, where $\beta$ changes from
  positive to negative at DM2 and the region inside DM1.
 \label{fig1}}
 \end{center}
\end{figure}

\begin{figure}
\centering
  \epsscale{1.40}\plotone{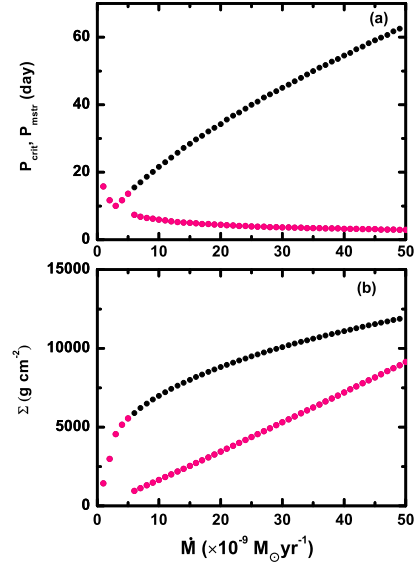}

 \caption{ The relationship between the gas density and star
 accretion rate. The upper panel (a) shows the orbital period at the location
 of maximum density with various star accretion rate.
 The black dot line represents the boundary of the dead zone and the active zone using equation
 (\ref{acrit}) while the purple line indicates the truncated location
 according to equation (\ref{mstr}). The bottom panel
 (b) displays the values of density at two maximum density locations.
 \label{fig2}}
  \vspace{0.4cm}
\end{figure}

According to equation (3), the accretion rate of star is
$\sim3.87\times 10^{-8}$$M_\odot~ \rm {yr^{-1}}$ for this system.
Nevertheless, the value will decrease, on average, with the
evolution of T Tauri star and its disk. Hence, herein we consider
the star accretion rate in the range of [$1\times 10^{-9}$, $5\times
10^{-8}$] $M_\odot~ \rm {yr^{-1}}$. The effective viscosity of the
disk is $\nu(a)=\alpha c_sh$, where $\alpha$ and $c_s$, represent
the efficiency factor of angular momentum transport, and sound speed
at the midplane, respectively; $h=c_s/\Omega$ means the isothermal
density scale height, $\Omega$ refers to the Kepler angular velocity
\citep{SS73}. Because of the effect of MRI, the values of $\alpha$
in the active zone and the dead zone are quite different. The
effective value of $\alpha$ is expressed as \citep{KL07,KL09}
\begin{equation}
\alpha_{\rm eff} (a)=\frac{\alpha_{\rm dead}-\alpha_{\rm mri}}{2}
\left[{\rm erf}\left(\frac{a-a_{\rm crit}}{0.1 a_{\rm
crit}}\right)+1\right]+\alpha_{\rm mri},
 \label{alpeff}
\end{equation}
where $\alpha_{\rm mri}$, $\alpha_{\rm dead}$ denote the value of
$\alpha$ in active zone and dead zone, respectively. Herein, we
choose $\alpha_{\rm dead}=0.001$ and $\alpha_{\rm mri}=0.01$
\citep{Sano}. The parameter $a_{\rm crit}$ in the error function erf
is the location of the boundary of MRI and $0.1 a_{\rm crit}$
represents the width of the transition zone. $a_{\rm crit}$ is
modeled as \citep{KL09}
\begin{eqnarray}
a_{\rm crit}=0.16 ~{\rm
AU}~\left(\frac{\dot{M}}{10^{-8}M_{\odot}~{\rm
yr}^{-1}}\right)^{4/9}
\left(\frac{M_{*}}{M_{\odot}}\right)^{1/3}
\nonumber\\
\times\left(\frac{\alpha_{\rm
mri}}{0.02}\right)^{-1/5}\left(\frac{\kappa_D}{1{\rm
cm^2g^{-1}}}\right),~~~~
 \label{acrit}
\end{eqnarray}
where $\kappa_D$ is the grain opacity.

Considering the disk depletion, the gas density profile can be
modified to be
\begin{equation}
\Sigma_g=\frac{\dot {M}}{3\pi\nu(a)}\exp\left(-\frac{t}{\tau_{\rm
dep}}\right),
\end{equation}
where $\tau_{\rm dep}$ refers to the disk depletion timescale, which
is observed as few million years \citep{hai01}. We adopt $\tau_{\rm
dep}=10^6$ yr in the simulations, where $t$ is the time of evolution.

Because of the stellar magnetic field, the gas disk is truncated at $a_{\rm mstr}$ \citep{konigl}
\begin{eqnarray}
a_{\rm mstr}=(1.06\times 10^{-2} ~{\rm AU}) \beta'
\left(\frac{R_*}{R_\odot}\right)^{12/7}\left(\frac{B_*}{1000{\rm
G}}\right)^{4/7}
\nonumber\\
\times\left(\frac{M_*}{M_\odot}\right)^{-1/7}\left(\frac{\dot{M}}{10^{-7}M_\odot~
{\rm yr^{-1}}}\right)^{-2/7}, \label{mstr}
\end{eqnarray}
where $R_*$, $R_\odot$, and $B_*$ refer to the radius of the star,
the radius of the sun and the magnetic field of the central star, respectively.
$\beta'\leq 1$ is a free parameter. Herein, we choose $\beta'$ = 1
corresponding to a typical Alfv$\dot {\rm e}$n radius in the way of
spherical accretion. Hereafter we use $P_{\rm crit}$ and $P_{\rm
mstr}$, the Keplerian orbital period, instead of $a_{\rm crit}$ and
$a_{\rm mstr}$ for convenience. Combining the effect of the magnetic
field, the gas density profile is substituted by
\begin{equation}
\Sigma_g=\frac{\dot {M}}{3\pi\nu(a)}\exp\left(-\frac{t}{\tau_{\rm
dep}}\right)\eta,\label{dens}
\end{equation}
where
\begin {equation}
\eta=0.5\left[{\rm erf}\left(\frac{a-a_{\rm mstr}}{0.1a_{\rm
mstr}}\right)+1\right],
\end{equation}
is induced by the truncation of the magnetic field.

Based on equation (\ref {dens}), we find that $\Sigma_g\propto
r^{-1}$ and there are generally two maxima in the gas density
profile. Figure \ref{fig1} shows the density profile with various
star accretion rates (see the top panel (a)). We label the inner
maximum density location as DM1, similarly, the outer one as DM2.
The locations of DM1 and DM2 change with the star accretion rate. In
Figure \ref{fig2} we show the density versus star accretion rate.
From Figure \ref{fig2} (the top panel (a)), we notice that, with a
decrease of star accretion rate, the values of DM1 and DM2 approach
each other until they merge into one, the combination occurs at
$\sim \dot M = 5\times 10^{-9}M_\odot ~{\rm yr^{-1}}$, corresponding
to a maximum density at an orbital period of 13 days. Herein we
choose $\alpha_{\rm dead}=0.001$, $\alpha_{\rm mri}=0.01$, and
$B_*=0.5~ {\rm KG}$, respectively, where the bottom panel (b)
displays the values of the density at DM1 and DM2, respectively.

\subsection{Planetary migration and eccentricity damping}
For the planets in KOI-152 system, their masses are estimated to be
less than 100 $M_\oplus$. From a classical planetary formation
theory, they may undergo type I or type II migration during their
evolution.

Type I migration is induced by angular-momentum exchange between gas
disk and planets. Based on linear analysis, the net momentum
loss on a planet causes an inward migration \citep{GT79,Ward97,Tan02}.
Under this assumption, the speed of type I migration is very fast.
In such situation, it is difficult to produce terrestrial planets.
Recently, there are several new theories on reducing the speed of type
I migration or even reverse the migrating direction \citep{BM08,KC08,KBK09,PP08,WZ11}.

Considering the uncertainty of type I migration, we adopt a
reduction of the migration speed, taking a timescale of type I
migration of an embryo with mass $m$ as \citep{Tan02}
\begin{eqnarray}
\tau_{\rm migI}=\frac{a}{|\dot{a}|}=\frac{1}{f_1}\tau_{\rm
linear}=\frac{1}{f_1(2.7+1.1\beta )}
\left(\frac{M_*}{m}\right)\left(\frac{M_*}{\Sigma_ga^2}\right)
\nonumber\\
\times\left(\frac{h}{a}\right)^2
\left[\frac{1+(\frac{er}{1.3h})^5}{1-(\frac{er}{1.1h})^4}\right]\Omega^{-1}\rm{yr},~~~~~~
\label{tauI}
\end{eqnarray}
where $e,~r,~h$, and $\Omega$ are eccentricity, distance from the
central star, scale hight of the disk, and the Keplerian angular
velocity, respectively. $\tau_{\rm Linear}$ is the timescale of
linear analysis result, $f_1$ is the reduction factor. Herein, we
choose $f_1$ = 0.03, 0.1, and 0.3, respectively, and $\beta =-d\ln
\Sigma_g/d\ln a$. $\Sigma_g$ means the gas density profile of the
disk expressed in equation (\ref{dens}). As the value of $\beta$ is
related to the density profile, the speed of type I migration may be
slowed down or even reversed at some special areas. In addition,
Figure 1 shows the values of $\beta$ using $\dot M$ = $2\times
10^{-8}~M_\odot ~{\rm yr^{-1}}$ as an example (see the bottom panel
(b)). According to equation (10), if $\beta<-2.45$, the timescale of
type I migration will transfer from positive to negative.
Furthermore, we notice that when embryos run across DM2 and DM1, the
migration speed will decrease, which may lead to the trapping of the
embryos there.

When it grows massive enough, a planet will start to experience type
II migration, as the strong torque caused by the planet will open a
gap in the gaseous disk \citep{lp93}. The timescale of type II
migration for a planet with mass $m$ is \citep{IL08}
\begin{eqnarray}
\tau_{\rm migII1}\simeq 5\times
10^5f_g^{-1}~~~~~~~~~~~~~~~~~~~~~~~~~~~~~~~~~~~~~~~~~~~~~~~~~~~~~~~
\nonumber\\
\times\left(\frac{C_2\alpha}{10^{-4}}\right)^{-1}\left(\frac{m}{M_J}\right)\left(\frac{a}{\rm{1AU}}\right)^{1/2}\left(\frac{M_*}{M_\odot}\right)^{-1/2}
\rm{yr},~~~~~~
\nonumber\\
\tau_{\rm migII2}=\frac{a}{|\dot{a}|}=0.7\times
10^5\left(\frac{\alpha}{10^{-3}}\right)^{-1}\left(\frac{a}{\rm
{1AU}}\right)\left(\frac{M_*}{M_\odot}\right)^{-1/2}\rm{yr},~~~~
\label{tpii}
\end{eqnarray}
where $\alpha$ is the efficiency factor of angular momentum
transport. When the mass of planet is comparable to that of the gas
disk, a fraction ($C_2 \sim 0.1$) of total angular momentum will
transfer between the planet and the disk in the evolution. In this
case, $\tau_{\rm migII1}$ applies. Herein we adopt
$\alpha=\alpha_{\rm dead}=0.001$, $C_2\alpha=10^{-4}$.  If the mass
of the gas disk is larger than that of the planet, it will migrate
with the gas disk over the timescale of $\tau_{\rm migII2}$. We
emphasize that the timescale of type II migration is larger than
that of type I.

A gap will form in the gas disk when the planet grows to massive
enough ($m>M_{\rm crit}$) \citep{IL08},
\begin {equation}
M_{\rm{crit}}\simeq
30\left(\frac{\alpha}{10^{-3}}\right)\left(\frac{a}{\rm
1AU}\right)^{1/2}\left(\frac{M_*}{M_\odot}\right)M_\oplus. \label
{cri}
\end {equation}
In this work, we assume that the planet will undergo type II
migration when its mass is greater than $M_{\rm crit}$.

Additionally, we consider the eccentricity damping induced by the
interactions between the gas disk and embryo \citep {GT79}. The
damping timescale for an embryo with mass $m$ is described as
\citep{Cre06}
\begin{eqnarray}
\tau_e=\left(\frac{e}{\dot{e}}\right)_{\rm
edamp}=\frac{Q_e}{0.78}\left(\frac{M_*}{m}\right)
\left(\frac{M_*}{a^2\Sigma_g}\right)\left(\frac{h}{r}\right)^4\Omega^{-1}
\nonumber\\
\times\left[1+\frac{1}{4}\left(e\frac{r}{h}\right)^3\right] {\rm
yr},~~~~~~~~~~~~~~~~~
\label{ede}
\end{eqnarray}
where $Q_e=0.1$ is a normalization factor to be consistent with
hydrodynamical simulations. The meanings of other symbols are
similar to those in equation (\ref{tauI}).

\section {NUMERICAL SIMULATIONS AND RESULTS}
To explore the secular evolution of the KOI-152 planetary system, we
assume that three planets had formed in the outer region of the
system and migrate toward the inner region due to the interactions
with the gas disk. Thus, the acceleration of a planet with mass
$m_i$ is given as
\begin{eqnarray}
\frac{d}{dt}\textbf{V}_i =
 -\frac{G(M_*+m_i
)}{{r_i}^2}\left(\frac{\textbf{r}_i}{r_i}\right) +\sum _{j\neq i}^N
Gm_j \left[\frac{(\textbf{r}_j-\textbf{r}_i
)}{|\textbf{r}_j-\textbf{r}_i|^3}- \frac{\textbf{r}_j}{r_j^3}\right]
\nonumber\\
+\textbf{F}_{\rm edamp}+\textbf{F}_{\rm
migI}({\rm or}+\textbf{F}_{\rm migII}),~~~~~~~~~ \label{eqf}
\end{eqnarray}
where $\textbf{r}_i$ and $\textbf{V}_i$ mean the position and
velocity vectors of the planet $m_i$ in the stellar-centric
coordinates, and the external forces are defined as
\begin {eqnarray}
\begin{array}{lll}
\textbf{F}_{\rm edamp} = -2\frac{\displaystyle (\textbf{V}_i \cdot
\textbf{r}_i)\textbf{r}_i}{\displaystyle r_i^2\tau_e},
\\
\cr\noalign{\vskip 0.5 mm} \textbf{F}_{\rm
migI}=-\frac{\displaystyle \textbf{V}_i}{\displaystyle 2\tau_{\rm
migI}},
\\
\cr\noalign{\vskip 0.5 mm} \textbf{F}_{\rm migII} =
-\frac{\displaystyle {\bf V}_i}{\displaystyle 2 \tau_{\rm migII}}.
\end{array}
\end{eqnarray}
Each planet is assumed to be initially in coplanar and near-circular
orbit and suffers from mutual gravitational interaction and the
effect exerted by the central star. The initial orbital elements of
each planet were randomly generated: argument of pericentre,
longitude of the ascending node, and mean anomaly were randomly set
between $0^{0}$ to $360^{0}$. In this way, we generate a set of 34
runs for simulation.

We numerically integrate the equations (\ref{eqf}) using a
time-symmetric Hermit scheme \citep{Aarseth}. If the distance from a
planet to the central star is smaller than the radii of the star we
originally set, we assume that the planet collides with the star.
Each run evolved for $10^8$ yr.

Based on equation (\ref{cri}), for $\alpha =10^{-3}$, if $a>0.125$
AU, then we have $M_{\rm crit}> 15~ M_\oplus$. Hence, if their
initial locations are all without 0.125 AU, KOI-152.02 and
KOI-152.03 will undergo type I migration, the outmost one will
experience type I or II migration. In this sense, two kinds of
models are assumed in the simulations: For Model 1, we suppose that
all planets will suffer type I migration; whereas for Model 2, we
simply consider that KOI-152.02 and KOI-152.03 will go through type
I migration but KOI-152.01 will undergo type II migration. According
to the aforementioned analysis, one of the planets in this system
may be captured at DM1 or DM2. In such case, taking into account the
range of the star accretion rate [$1\times 10^{-9}$, $5\times
10^{-8}$] $M_\odot~ \rm {yr^{-1}}$, we find that the positions of
DM2 are at $\sim$ 13, 26, and 48 days, respectively, in three groups
for Model 1. In addition, the conditions of density profile of
various groups and models are summarized as follows.

\textbf{Model 1}

Group 1: For $\dot M = 5\times 10^{-8}~ M_{\odot}~{\rm yr^{-1}}$,
the density profile $\Sigma_g$ is labeled as the blue line in Figure
\ref{fig1}.  DM2 = 48.039 days, KOI-152.01 is likely to be captured
at $a_{\rm crit}$.

Group 2: For $\dot M = 2\times 10^{-8}~ M_{\odot}~{\rm yr^{-1}}$,
$\Sigma_g$ is labeled as the black line (Figure \ref{fig1}). DM2 =
26.08 days, KOI-152.02 is likely to be captured at $a_{\rm crit}$.

Group 3: For $\dot M = 1\times 10^{-9}~ M_{\odot}~{\rm yr^{-1}}$,
$\Sigma_g$ is labeled as the green line (Figure \ref{fig1}). DM2 =
13.067 days, KOI-152.03 is likely to be captured at $a_{\rm crit}$.

Group 4: For $\dot M = 1\times 10^{-9}~ M_{\odot}~{\rm yr^{-1}}$,
$\Sigma_g$ is labeled as the red line (Figure \ref{fig1}). DM2 =
51.9 days, KOI-152.01 is likely to be captured at $a_{\rm crit}$.

\textbf{Model 2}

Comparison with Group 2 of Model 1: For $\dot M = 2\times 10^{-8}~
M_{\odot}~{\rm yr^{-1}}$, $\Sigma_g$ labeled as the black line in
Figure \ref{fig1}. DM2 = 26.08 days, KOI-152.02 is likely to be
captured at $a_{\rm crit}$. However, KOI-152.01 will undergo type II
migration.

\begin{figure}
\begin{center}
  \epsscale{1.20}\plotone{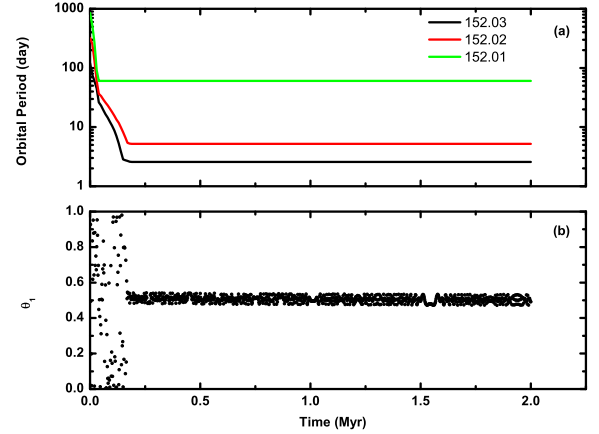}
 \caption{The evolution of orbital periods and resonant angle in
Group 1. At $\sim$ 0.2 Myr, KOI-152.02 and KOI-152.03 are evolved
into a 2:1 MMR in the evolution. The upper panel (a) shows the time
evolution of the orbital periods. Planet 03 halts at DM2 first until Planet 02 comes. Planet 02 kicks 03 into the
inner region when it migrates towards DM2 and the outmost planet keeps
Planet 02 repeating this process just as the first two planets do.
At the end of the run, Planet 03 stops at DM1. The bottom panel (b)
shows one of the 2:1 resonant angles of Planet 02 and 03.
(Hereafter, we label
$\theta_1=2\lambda_{02}-\lambda_{03}-\varpi_{02},~\theta_2=2\lambda_{02}-\lambda_{03}-\varpi_{03},~\theta_3=2\lambda_{01}-\lambda_{02}-\varpi_{01},~\theta_4=2\lambda_{01}-\lambda_{02}-\varpi_{02}.$)
 \label{fig3}}
 \end{center}
\end{figure}

\begin{figure*}
\begin{center}
 \plotone{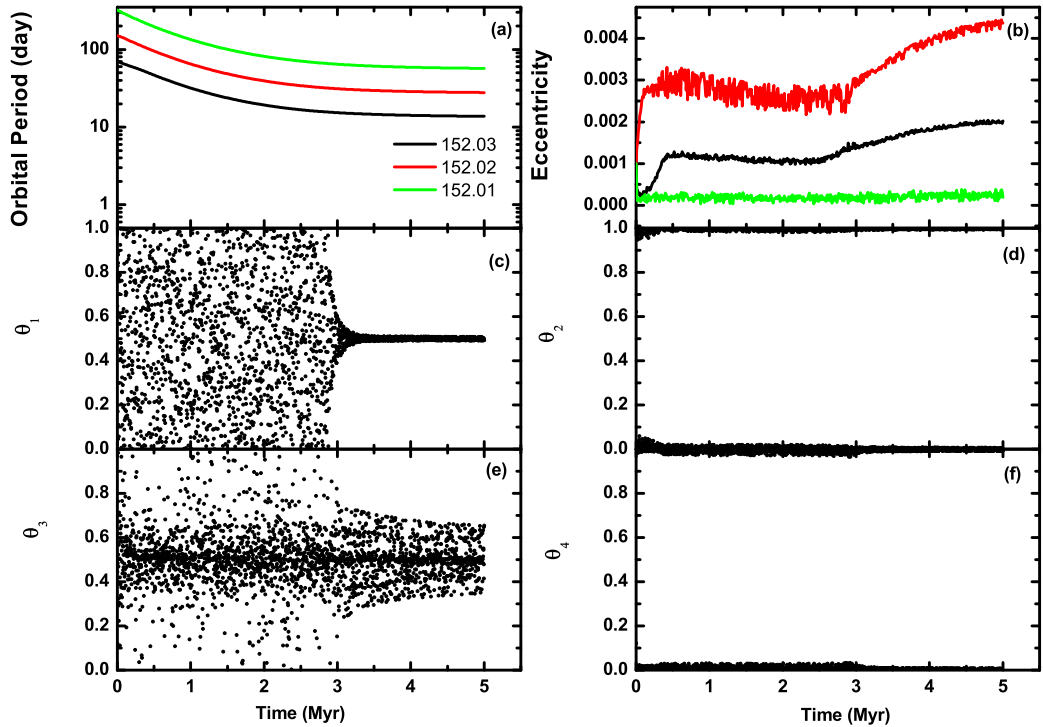}
 \caption{ The results of the evolution for Group 3-2. The two upper panels (a and b) show the
evolution of orbital periods and eccentricities. The middle and
bottom panels (c-f) are indicative of the resonant angles of three
planets. In this run, they are captured into 4:2:1 MMR over very
shorter timescale. In the end of this run, a planetary configuration
similar to KOI-152, which consists of three planets, is formed.
 \label{fig4}}
 \end{center}
\end{figure*}

\begin{figure*}
\begin{center}
 \plotone{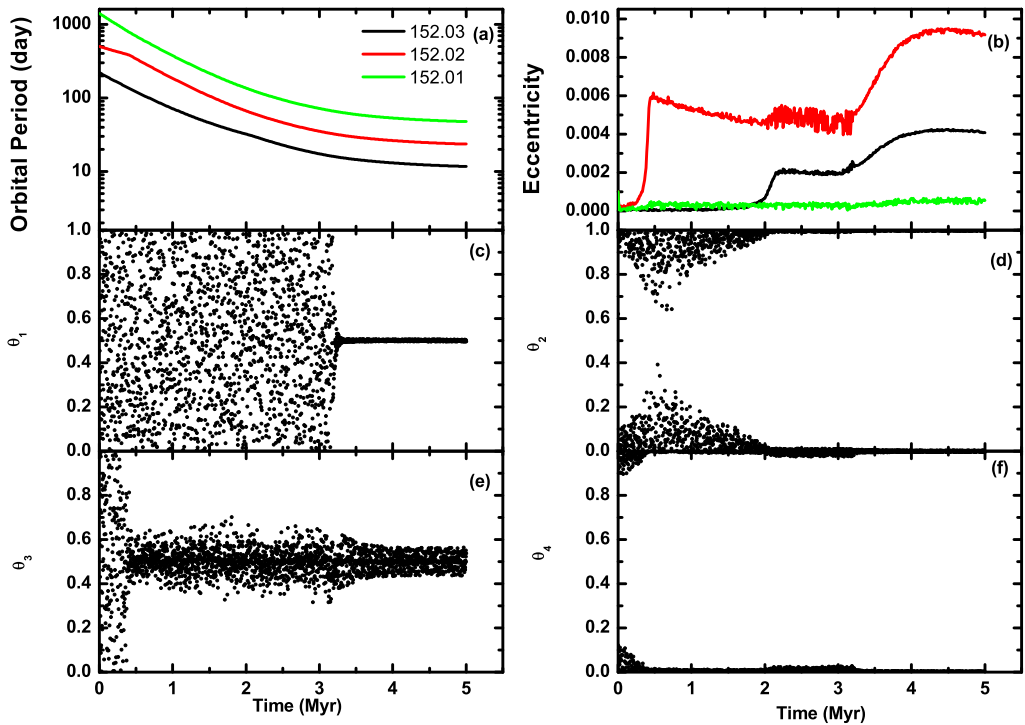}
 \caption{The results of the evolution Group 3-3. The two upper panels (a and b) show the
evolution of orbital periods and eccentricities. The middle and
bottom panels (c-f) exhibit their resonant angles. In this run, they
are captured into 4:2:1 MMR within a shorter timescale. A planetary
configuration analogous to KOI-152, consisting of three planets, is
generated.
 \label{fig5}}
 \end{center}
\end{figure*}

\begin{table*}
\centering \caption{Detailed information of five groups in the
simulations. Terminal periods are the orbital periods of three
planets at the end of run. $P_{\rm crit}$ and $P_{\rm mstr}$
represent Keplerian orbital periods in association with the regime
of DM1 and DM2. \vspace{0.5cm}
 \label{tb2}}
\begin{tabular*}{16cm}{@{\extracolsep{\fill}}lllllll}
\tableline
  ID & $\dot M$  & Initial periods &$f_1$      & Terminal periods& Terminal periods ratios  &$P_{\rm{crit}}$, $P_{\rm{mstr}}$\\
& $\rm ({M_\odot~ yr^{-1}})$ & (days)&& (days)& ($P_{01}/P_{02}$, $P_{02}/P_{03}$)& (days)\\
\tableline
Group 1&$5\times 10^{-8}$& 120, 320, 850 & 0.1 &   2.58, 5.2, 60.5&11.63, 2.02& 48.039, 2.443\\
Group 2-1&$2\times 10^{-8}$& 120, 320, 850 & 0.03 &   3.8, 7.75, 32.83&4.24, 2.04&26.08, 3.618 \\
Group 2-2&$2\times 10^{-8}$& 120, 320, 850 & 0.1 &   1.899, 12.19, 24 &1.97, 6.42&26.08, 3.618 \\
Group 2-3&$2\times 10^{-8}$& 120, 320, 850 & 0.3 &   3.8, 7.67, 32.85 &4.28, 2.02&26.08, 3.618 \\
Group 2-4&$2\times 10^{-8}$& 120, 320, 20012 & 0.03 &   31.72, 64.23, 6960.77 &108.37,2.02&26.08, 3.618 \\
Group 2-5&$2\times 10^{-8}$& 120, 320, 20012 & 0.1 &   3.82, 7.7, 32.85 &4.27, 2.01&26.08, 3.618 \\
Group 2-6&$2\times 10^{-8}$& 120, 320, 20012 & 0.3 &   3.78, 5.72, 32.83 &5.74, 1.51&26.08, 3.618 \\
Group 3-1&$1\times 10^{-9}$& 120, 320, 850 & 0.1 &   3.5, 7.12, 14.73&2.07, 2.03& 13.067\\
Group 3-2&$1\times 10^{-9}$& 70, 150, 320 & 0.03 &   13.78, 27.87, 57.07 &2.05, 2.02&13.067\\
Group 3-3&$1\times 10^{-9}$& 220, 500, 1400 & 0.1 &   11.71, 23.56, 47.64&2.02, 2.01& 13.067\\
Group 4&$1\times 10^{-9}$& 220, 320, 850 & 0.1 &   14.17, 28.56, 58.54 &2.05, 2.01&51.9\\
KOI-152&&&&13.48, 27.40, 52.09&1.90, 2.03&\\
 \tableline
\end{tabular*}
\end{table*}

\subsection{Model 1: All planets undergo type I migration}

In order to understand the influence of the star accretion rate and
the speed of type I migration, we perform four groups of
simulations. Table 2 are listed the detailed information for the
dominatant results of each group, where the last column shows the
Keplerian orbital periods at the locations of DM1 and DM2.
\begin{figure}
\begin{center}
  \epsscale{1.3}\plotone{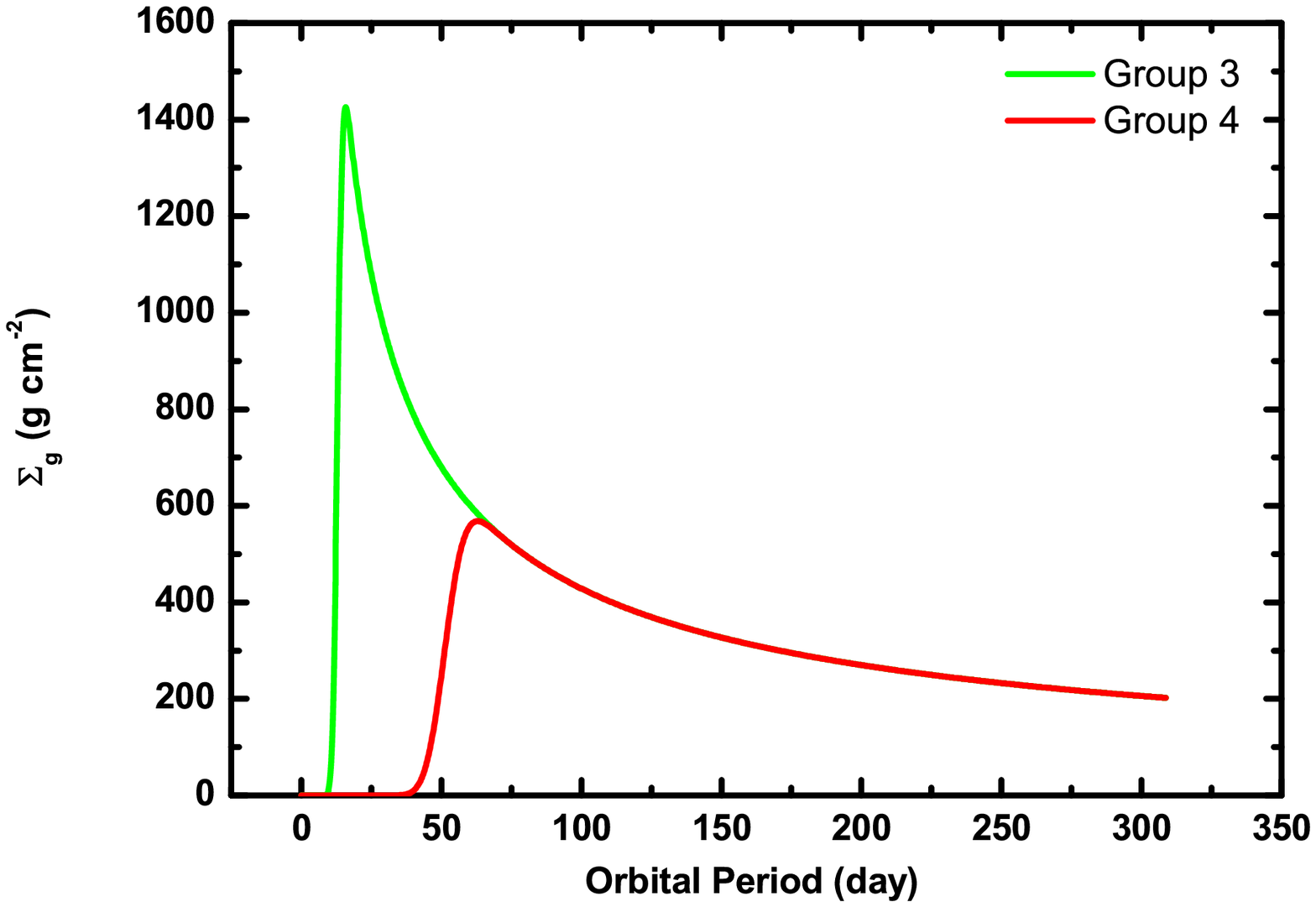}
 \caption{The gas density profile in Group 3 and
4. Because of a high magnetic field of the star, the gas disk will
be truncated at distant region with low gas density in Group 4.
 \label{fig6}}
 \end{center}
\end{figure}

\subsubsection{Group 1: KOI-152.01 captured at
$a_{\rm{crit}}$}

In this group, DM1 and DM2 occur at $\sim$ 2.4 days and 48 days,
separately, with $\dot M = 5\times 10^{-8} M_{\odot}~{\rm yr^{-1}}$.
From the profile of gas density, we note that, if Planet 01 is
likely to be trapped at DM2, the resulting configuration may be
analogous to KOI-152. In total, we performed five runs to examine
how the planets come into the resonant region, where Table 2 reports
typical outcomes in our simulations. For example, in a typical run,
Planet 02 or 03 will continue to move inward until reaching DM1 in
cases where they have been kicked inside of DM2. Subsequently,
Planet 01 is trapped at $\sim$ 60 days as shown in Figure 3 and the
two inner planets are eventually captured into 2:1 MMR. From Figure
1, we show that if a planet is pushed into the region inside DM2,
the value of $\beta$ changes from negative to positive before it
approaches DM1. The change in sign of $\beta$ is the reason that the
two inner bodies migrate towards DM1. However, no matter how we
varied the speed of type I migration or re-scaled the initial
position of Planet 01, (e.g., much more distant than the value given
in Table 2), Planet 02 and 03 still rush into the inner zone about
DM1 in the evolution. Hence, we may safely conclude that a
configuration like KOI-152 cannot be generated when the star is too
young with high star accretion rate. To sum up the above outcomes,
for five runs, we find that Planet 02 and 03 are locked into 2:1 MMR
in three cases during dynamical evolution, whereas they are captured
into 3:2 MMR in the other two runs.

In this scenario, a planetary configuration is finally created, that
consists of two planets trapped into a 2:1 MMR or 3:2 MMR at DM1 and
the outmost one resided inside DM2, far away from DM1.

\begin{figure*}
\begin{center}
  \epsscale{1} \plotone{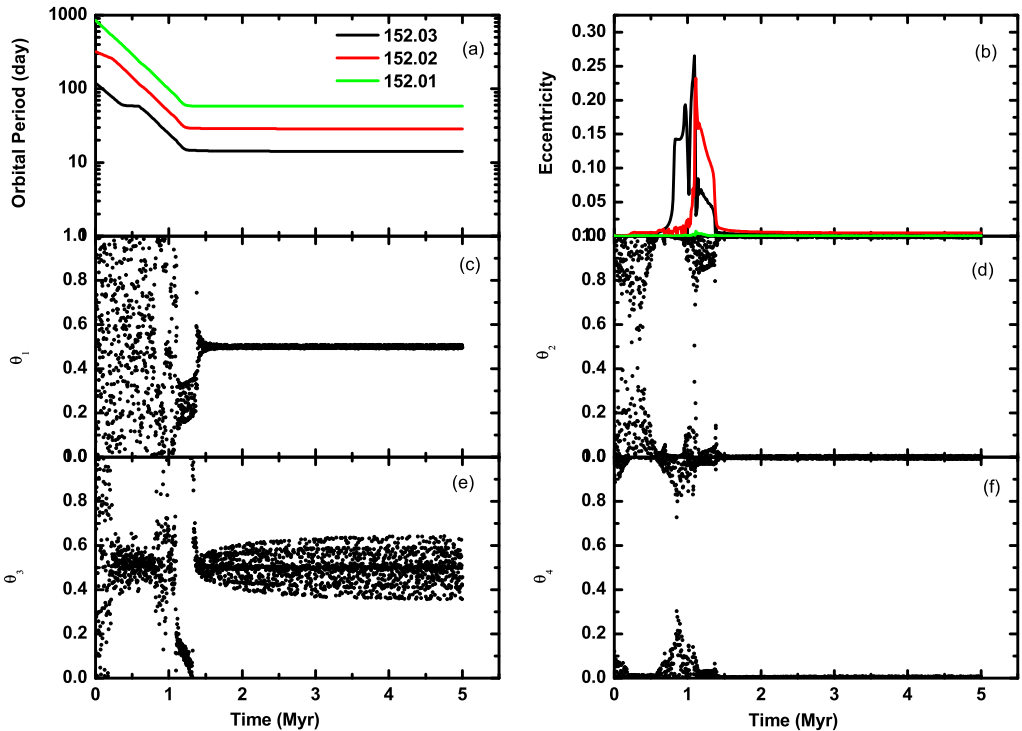}
 \vspace{0cm}
 \caption{The results of the evolution for Group 4. The magnetic field
 is $B_*=2.5$ KG. Note that three planets are captured into 4:2:1 MMR.
 The middle and bottom panels (c-f) show the time variations of their resonant angles.
 \label{fig7}}
 \end{center}
\end{figure*}

\subsubsection{Group 2: KOI-152.02 captured at $a_{\rm{crit}}$}

By adopting $\dot M=2\times 10^{-8}~M_\odot~\rm{ yr^{-1}}$, we have
DM1 and DM2 located at $\sim 3.6,~26$ days, respectively.  In this
group, we performed six runs in total. The results of all runs are
reported in Table \ref{tb2}. From Table \ref{tb2}, we observe that,
similar to the cases of Group 1, Planet 03 is quickly captured at
DM1 as it is thrown into the region inside DM2. In Run 1-3, we
utilize the same initial conditions but choose a variational
migration speed. In these runs, Planet 02 cannot stay at DM2 but is
kicked into the inner region by the perturbation of Planet 01, then
it falls into MMR with Planet 01 (run 2) or Planet 03 (run 1, 3). In
Run 4-6, Planet 01 is initially well separated from the other two,
by placing the outer planet in a starting location of much more
distant from the central star. The results of Run 5 and 6 are
analogous to that of Run 1-3. The variation of $\beta$ is the
governing reason for the evolution of Run 1-3 and 5-6.  However,
owing to slow migration speed, Run 4 differs from other runs, where
three planets cannot pass through DM2. Here, one may notice that it
is impossible for three planets to form a configuration resembling
to KOI-152. From the analysis of six runs, we summarize the
simulation outcomes, where Planet 02 and 03 are in a 2:1 MMR for
four runs and in one run they are captured into a 3:2 MMR , whereas
in another special case Planet 01 and 02 become trapped into a 2:1
MMR over the evolution.

The created configuration in Group 2 varies with the value of $f_1$.
The inner planet is always trapped in the boundary of the inner hole
unless they undergo a lower speed of type I migration and the
outmost planet is formed farther simultaneously. Thus, in such
scenario, a configuration similar to KOI-152 cannot be created.

\subsubsection{Group 3: KOI-152.03 captured at $a_{\rm{crit}}$}

In the simulations, we assume the star accretion rate is equal to
$1\times 10^{-9}~M_\odot~ \rm{yr^{-1}}$. From the density profile
shown in Figure \ref{fig1}, we see that DM1 and DM2 tend to combine
into one position at $\sim$ 13 days. In this group, we totally
carried out 10 runs. Details of three runs are also shown in Table
\ref{tb2}. If we examine the type I migration considering $f_1\geq
0.1$, the results of evolution then bear resemblance to the case of
Group 3-1, where KOI-152.01 is trapped at $\sim$ 14 days in the
meantime the other two planets jump into the inner region at $P<10$
days, unless we started them at distant orbits from the star or slow
down the speed of type I migration. In Group 3-2, we lower the
migration speed and the result is shown in Figure \ref{fig4}, where
Planet 03 is capable to being trapped at $\sim$ 14 days. The
simulation results are consistent with the current observational
values of KOI-152. When Planet 02 and 01 keep pace with Planet 03,
two pairs are all captured into 2:1 MMRs. Group 3-3 simulates the
condition related to a high speed of type I migration and distant
initial orbits. Figure \ref{fig5} displays that Planet 01 and 02 are
locked into 2:1 MMR in the dynamical evolution as well as Planet 02
and 03. In conclusion, according to our evaluation, three planets
are eventually captured into Laplace resonance, with the resonant
angles in a libration about 180$^\circ$ for Group 3-2 and 3-3.  When
the planets are trapped into MMRs, their eccentricities are excited
during the evolution; but due to strong gas damping, they cannot be
pumped up to large values.

Taking $f_1\leq 0.1$ into account, a configuration similar to
KOI-152 is formed, where two pairs of planets are in the 2:1 MMRs.

\subsubsection{Group 4: High magnetic field star}

Herein we consider that the star bears a high magnetic field of
$B_*=2.5$ KG. In comparison with Group 3, for a high magnetic field
star, the truncation of the inner hole is at $\sim$ 50 days, farther
than that a low magnetic field star as shown in Figure 6. Herein we
carried out three runs. Figure 7 illustrates the results of a
typical run, where the middle and bottom panels (c-f) show the
resonant angles for the planets. From the figure, we note that three
planets are captured into a 4:2:1 MMR at $\sim$ 1 Myr.
Simultaneously, we examine the resonant angles and find that three
planets are also locked into a Laplacian resonance, librating about
0$^\circ$ at $\sim$ 4 Myr. When they are captured into MMR, Planet
02 and 03 may migrate towards the inner region of the hole with
little gas. Consequently, the influence of the gas damping is not
dominant. Thus, the eccentricities of Planet 02 and 03 are stirred
up to $\sim$ 0.2 when they are trapped into a 2:1 MMR. However,
Planet 01 remains outside the hole, and its eccentricity finally
decreases to $\sim$ 0.002.  For one paradigm, because of fast speed
of type I migration, the system is totally destroyed. In another
run, the produced planetary configuration is quite analogous to the
typical Laplacian geometry,  with the resonant angles librating at
180$^\circ$. For $f_1\leq 0.1$, we have a planetary configuration
similar to KOI-152.

\subsection{Model 2: KOI-152.01 undergoes type II migration}

\begin{figure}
\begin{center}
   \epsscale{1.2}\plotone{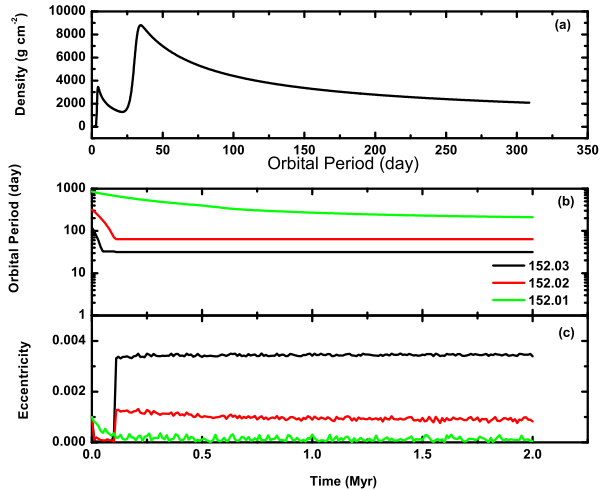}
 \vspace{0cm}
 \caption{The orbital evolution for Model 2. Two inner planets undergo type I migration
 while the outer one suffers from type II migration.
 The upper panel (a) shows the density profile, and the middle and bottom panels (b and c)
 exhibit the time evolution of the semimajor axes and eccentricities, respectively.
 \label{fig8}}
 \end{center}
\end{figure}

According to our above analysis, Planet 01 may perform type II
migration. In this model, we set Planet 02 and 03 to undergo type I
migration but Planet 01 suffers from type II migration. We carry out
10 runs for Model 2, by varying the initial position of the outmost
planet. But owing to a slow speed of type II migration, Planet 01
cannot reach its nominal position even if it is placed at the orbit
with a period of 70 days in the beginning. Figure \ref{fig8} shows a
typical run of the simulations. In this run, the density profile is
the same with that in model 1 Group 2, $f_1$=0.1 for Planet 02 and
03. In the end, the simulation results show that three planets
migrate to 31.72, 63.86 and 211.13 days, respectively. However,
Planet 01 is a little deviated from the present-day observation.

Under this circumstance, Planet 01 cannot approach its estimated
position regardless the initial location of the outmost planet. In
addition, the inner two planets are always trapped into 2:1 MMR at
DM2. In a word, we cannot generate a Laplacian configuration for
three planets from Model 2.

\begin{table*}
\centering \caption{The orbital periods of each planetary candidate
and the ratios between them. $P_{01}$, $P_{02}$,$P_{03}$, and
$P_{04}$ are, respectively, the orbital periods of the planets in
the systems. \vspace{0.5cm}
 \label{tb3}}
\begin{tabular*}{12cm}{@{\extracolsep{\fill}}llllllll}
\tableline
  ID & $P_{01}$  & $P_{02}$ &$P_{03}$&$P_{04}$& $P_{01}/P_{02}$& $P_{02}/P_{03}$&$P_{03}/P_{04}$\\
KOI& (days)& (days)&(days) &(days) & &\\
\tableline
152&52.09119 & 27.40415& 13.484 &- &1.9009 &2.0323&-\\
571&13.34331 & 7.26733 &3.886758& -&1.8361 &1.8698&-\\
665&5.867973 & 3.07154 &1.611912& -&1.9104 &1.9055&-\\
733&11.34917 &5.924992 &3.132968& -&1.9155 &1.8912&-\\
829&38.5596  &18.64902 &9.75222 & -&2.0676 &1.9123&-\\
898&20.08923 &9.77059  &5.16991 & -&2.0561 &1.8899&-\\
899&15.36813 &7.11388  &3.306569& -&2.1603 &2.1514&-\\
1426&150.0341&74.91443 &38.87641& -&2.0027 &1.9270&-\\
1860&12.2094&6.3194&3.0765&-&1.9321&2.0541&-\\
1895&32.1349&17.2812&8.4575&-&1.8595&2.0433&-\\
935&87.6464  &42.6329  &20.85987& 9.6168&2.0558 &2.0438&2.1691\\
148&42.89554 &9.67374  &4.777978& -&4.4342 &2.0247&-\\
\tableline
\end{tabular*}
\end{table*}

\section{Conclusions and Discussions}
In this work, we have extensively investigated the formation of
configuration for the KOI-152 system using numerical simulations. We
assume three planets formed in a region far away from their current
locations. The embedded planets in the gas disk will migrate into
the regime much closer to the central star. In order to produce a
configuration similar to KOI-152, there are some requirements for
the disk and the star. In summary, we reach the following
conclusions:

1. If KOI-152.01 survives as predicted, no type II migration
happened for three planets in the system.

2. From equation (12), the masses of three planets can be limited to
be $m_{03}\leq 15,~m_{02}\leq19$, and ~$m_{01}\leq24~M_\oplus$,
respectively.  According to core-accretion model, a solid core may
continue to accrete gas envelop until it grows up into 2-10
$M_\oplus$ \citep{Barnes09,Boden,Hub,Ikoma01,pollack}, therefore
simply judging from the radii of the planets and the formation
scenario, we cannot clearly identify the composition of the planets,
and their masses are not well determined. Based on the gap opening
criteria, the masses of three planets are evaluated to be (9-15)
$M_\oplus$, (9-19) $M_\oplus$, and (20-24) $M_\oplus$, respectively.

3. Low star accretion rate and high magnetic field of the star are
propitious to the formation of the configuration of KOI-152. Low
star accretion rate may imply that the formation of three planets
takes place at the later stage of the star evolution, where DM1 and
DM2 are merged into one. From Figure \ref{fig2}, we learn that the
boundary is about $\dot M = 5\times 10^{-9} { M_{\odot}~\rm
yr^{-1}}$. In this sense, one requirement for the formation of
KOI-152 is $\dot M \leq 5\times 10^{-9} M_{\odot}~{\rm yr^{-1}}$.

4. Low speed of type I migration with $f_1\leq 0.1$ facilitates the
formation of the KOI-152 system. This conclusion agrees with those
mentioned by  Ida \& Lin (2008) and Wang \& Zhou (2011).

5. In this work, we had not examined tidal interactions between
planets and central star after planetary formation scenario. On the
basis of tidal timescale \citep{ML04,ZL08}, the eccentricities of
Planet 01, 02, 03, with their nominal orbital elements will be
damped in $\tau_{\rm tide}$, $1.3\times 10^9Q'$, $2\times 10^8Q'$,
and $9.6\times 10^6Q'$ years, respectively,  where $Q'$ is the tidal
dissipation factor and we adopt $\rho~=~3~\rm{g~cm^{-3}}$ for them.
Meanwhile, the semimajor axes are also decreasing over the
timescales of $\tau_{\rm tide}/(2e^2)$. In that case, the orbits of
the planets will become a little closer to the star that those
without consideration of tidal effects. Owing to shorter timescales,
the inner planets migrate faster, driving all of them out of MMR.
Then, the configuration of KOI-152 forms and three planets are in
near MMRs. This scenario may be suitable to account for the
formation of other systems with planetary configurations like
KOI-152.

Among 16 months data of Kepler, 242 target stars host two planet
candidates, 85 with three, 25 with four planets, eight with five
planets and one with six \citep{Fab12}. We examine all candidates
and find that ten systems may have two pairs of the planets involved
in near 2:1 MMRs, e.g., a near Laplacian resonance configuration. The orbital
periods and their ratios are shown in Table
3\footnote[1]{http://archive.stsci.edu}. The ratios of orbital
periods are all in the range of [1.83, 2.18], which may imply that
they are likely to bear a formation scenario similar to KOI-152. In
addition, the configuration of KOI-148 is close to the outcomes of
Group 2-1 and 2-3 where two planets are simply trapped in 2:1 MMR,
which demonstrates that the system may be formed when the star is
older than the case of KOI-152 with ${\dot M}=2\times 10^8~{\rm
M_{\odot}~yr^{-1}}$.

\citet{Marcy01} revealed that there were two giant planets involved
in a 2:1 MMR orbiting the star GJ 876. However, after the fourth
planet was discovered in the GJ 876 system, the previously known 2:1
MMR configuration then becomes a Laplacian resonance configuration,
where three planets are trapped into a 4:2:1 MMR with the resonant
angles librating at 0$^\circ$, respectively, differing from the
Galilean moons of Jupiter in a libration at 180$^\circ$
\citep{Rivera2010}. In such configurations, a planetary system will
remain stable at least one billion years. A great many of previous
works had investigated the dynamics and stability of the system
consisting of three planets at then
\citep{Ji02,Ji03,Zhou05,Zhang10}. From the former results, we learn
that the migration scenario may be responsible for the formation of
two giants which is locked into 2:1 MMR. Thus under the same
formation scenario for KOI-152, three planets may be also trapped
into a 4:2:1 MMR, showing a resemblance to those of GJ 876. Hence,
we may reach a safe conclusion that the near Laplacian
configurations are quite common in the planetary systems as also
revealed by Kepler and our work may provide some substantial clues
to the formation of such intriguing systems.

\acknowledgments{We thank the anonymous referee for good comments
that helped to improve the contents of the manuscript. W.S. is
supported by NSFC (Grants No. 10925313, 10833001) and China
Postdoctoral Science Foundation (Grant No. 2011M500962). J.J.H.
acknowledges the support by the National Natural Science Foundation
of China (Grants No. 10973044, 10833001), the Natural Science
Foundation of Jiangsu Province (Grant No. BK2009341), the Foundation
of Minor Planets of Purple Mountain Observatory, and the innovative
and interdisciplinary program by CAS (Grant No. KJZD-EW-Z001).}


\begin{thebibliography}{}

\bibitem[\protect\citeauthoryear{Aarseth}{2003}]{Aarseth}
Aarseth, S.~J.\ 2003, Gravitational N-Body Simulations, by Sverre
J.~Aarseth, pp.~430.~ISBN 0521432723.~Cambridge, UK: Cambridge
University Press, November 2003.

\bibitem[\protect\citeauthoryear{Barnes et al.}{2009}]{Barnes09}
Barnes, R., Jackson, B., Raymond, S.~N., West, A.~A., \& Greenberg, R.\ 2009, \apj, 695, 1006

\bibitem[\protect\citeauthoryear{Baruteau \& Masset}{2008}]{BM08}
Baruteau, C., \& Masset, F.\ 2008, \apj, 672, 1054

\bibitem[\protect\citeauthoryear{Batalha et al.}{2012}]{Bata12}
Batalha, N.~M., Rowe, J.~F., Bryson, S.~T., et al.\ 2012, arXiv:1202.5852

\bibitem[\protect\citeauthoryear{Bodenheimer et al.}{2000}]{Boden}
Bodenheimer, P., Hubickyj, O., \& Lissauer, J.~J.\ 2000, Icarus,
143, 2

\bibitem[\protect\citeauthoryear{Borucki et al.}{2010}]{Borucki10}
Borucki, W.~J., Koch, D., Basri, G., et al.\ 2010, Science, 327, 977

\bibitem[\protect\citeauthoryear{Borucki et al.}{2011}]{Borucki11}
Borucki, W.~J., Koch, D.~G., Basri, G., et al.\ 2011, \apj, 736, 19

\bibitem[\protect\citeauthoryear{Cresswell \& Nelson}{2006}]{Cre06}
Cresswell, P., \& Nelson, R.~P.\ 2006, \aap, 450, 833

\bibitem[\protect\citeauthoryear{Fabrycky et al.}{2012}]{Fab12}
Fabrycky, D.~C., Lissauer, J.~J., Ragozzine, D., et al.\ 2012, arXiv:1202.6328

\bibitem[\protect\citeauthoryear{Ford et al.}{2011}]{Ford11}
Ford, E.~B., Rowe, J.~F., Fabrycky, D.~C., et al.\ 2011, \apjs, 197, 2


\bibitem[\protect\citeauthoryear{Goldreich \& Tremaine}{1979}]{GT79}
Goldreich, P., \& Tremaine, S.\ 1979, \apj, 233, 857

\bibitem[\protect\citeauthoryear{Goldreich \& Tremaine}{1980}]{GT80}
Goldreich, P., \& Tremaine, S.\ 1980, \apj, 241, 425

\bibitem[\protect\citeauthoryear{Haisch et al}{2001}]{hai01}
Haisch, K.~E., Jr., Lada, E.~A., \& Lada, C.~J.\ 2001, \apjl, 553, L153


\bibitem[\protect\citeauthoryear{Hubickyj et al.}{2005}]{Hub}
Hubickyj, O., Bodenheimer, P., \& Lissauer, J.~J.\ 2005, Icarus,
179, 415

\bibitem[\protect\citeauthoryear{Ida \& Lin}{2004}]{idalin04}
Ida, S., \& Lin, D.~N.~C.\ 2004, \apj, 604, 388

\bibitem[\protect\citeauthoryear{Ida \& Lin}{2008}]{IL08}
Ida, S., \& Lin, D.~N.~C.\ 2008, \apj, 673, 487

\bibitem[\protect\citeauthoryear{Ikoma et al.}{2001}]{Ikoma01}
Ikoma, M., Emori, H., \& Nakazawa, K.\ 2001, \apj, 553, 999

\bibitem[\protect\citeauthoryear{Ji et al.}{2002}]{Ji02}
Ji, J., Li, G., \& Liu, L.\ 2002, \apj, 572, 1041

\bibitem[\protect\citeauthoryear{Ji et al.}{2003}]{Ji03}
Ji, J., Liu, L., Kinoshita, H., et al.\ 2003, \apjl, 591, L57

\bibitem[\protect\citeauthoryear{Ji et al.}{2011}]{Ji11}
Ji, J., Jin, S., \& Tinney, C.~G.\ 2011, \apjl, 727, L5

\bibitem[\protect\citeauthoryear{Koenigl}{1991}]{konigl}
Koenigl, A.\ 1991, \apjl, 370, L39

\bibitem[\protect\citeauthoryear{Kley \& Crida}{2008}]{KC08}
Kley, W., \& Crida, A.\ 2008, \aap, 487, L9

\bibitem[\protect\citeauthoryear{Kley et al.}{2009}]{KBK09}
Kley, W., Bitsch, B., \& Klahr, H.\ 2009, \aap, 506, 971

\bibitem[\protect\citeauthoryear{Kokubo \& Ida}{2002}]{KI02}
Kokubo, E., \& Ida, S.\ 2002, \apj, 581, 666

\bibitem[\protect\citeauthoryear{Kretke \& Lin}{2007}]{KL07}
Kretke, K.~A., \& Lin, D.~N.~C.\ 2007, \apjl, 664, L55

\bibitem[\protect\citeauthoryear{Kretke et al.}{2009}]{KL09}
Kretke, K.~A., Lin, D.~N.~C., Garaud, P., \& Turner, N.~J.\ 2009, \apj, 690, 407

\bibitem[\protect\citeauthoryear{Lee \& Peale}{2002}]{lee02}
Lee M. H., \& Peale S. J. 2002, \apj, 567, 596

\bibitem[\protect\citeauthoryear{Lin \& Papaloizou}{1993}]{lp93}
Lin, D. N. C, \& Papaloizou, J. C. B. 1993, in: E.H. Levy \& J.I. Lunine (eds.), Protostars and Planets III, (Tucson: Unv. Arizona)

\bibitem[\protect\citeauthoryear{Lin et al.}{1996}]{lin96}
Lin, D.~N.~C., Bodenheimer, P., \& Richardson, D.~C.\ 1996, \nat, 380, 606

\bibitem[\protect\citeauthoryear{Lissauer et al.}{2011}]{Lissauer11}
Lissauer, J.~J., Ragozzine, D., Fabrycky, D.~C., et al.\ 2011, \apjs, 197, 8

\bibitem[\protect\citeauthoryear{Masset \& Snellgrove}{2001}]{MS01}
Masset, F., \& Snellgrove, M.\ 2001, \mnras, 320, L55

\bibitem[\protect\citeauthoryear{Marcy et al.}{2001}]{Marcy01}
Marcy, G.~W., Butler, R.~P., Fischer, D., et al.\ 2001, \apj, 556, 296

\bibitem[\protect\citeauthoryear{Mardling \& Lin}{2004}]{ML04}
Mardling, R.~A., \& Lin, D.~N.~C.\ 2004, \apj, 614, 955

\bibitem[\protect\citeauthoryear{Natta et al.}{2006}]{natta}
Natta, A., Testi, L., \& Randich, S.\ 2006, \aap, 452, 245

\bibitem[\protect\citeauthoryear{Paardekooper \& Papaloizou}{2008}]{PP08}
Paardekooper, S.-J., \& Papaloizou, J.~C.~B.\ 2008, \aap, 485, 877

\bibitem[\protect\citeauthoryear{Papaloizou \& Terquem}{2010}]{PT10}
Papaloizou, J.~C.~B., \& Terquem, C.\ 2010, \mnras, 405, 573

\bibitem[\protect\citeauthoryear{Pollack et al.}{1996}]{pollack}
Pollack, J.~B.,Hubickyj, O., Bodenheimer, P., et al.\ 1996, Icarus,
124, 62

\bibitem[\protect\citeauthoryear{Pringle}{1981}]{pringle81}
Pringle, J.~E.\ 1981, \araa, 19, 137

\bibitem[\protect\citeauthoryear{Rasio \& Ford}{1996}]{RF96}
Rasio, F.~A., \& Ford, E.~B.\ 1996, Science, 274, 954

\bibitem[\protect\citeauthoryear{Rivera et al.}{2010}]{Rivera2010}
Rivera, E.~J., Laughlin, G., Butler, R.~P., et al.\ 2010, \apj, 719, 890

\bibitem[\protect\citeauthoryear{Sano et al.}{2000}]{Sano}
Sano, T., Miyama, S.~M., Umebayashi, T., \& Nakano, T.\ 2000, \apj, 543, 486

\bibitem[\protect\citeauthoryear{Shakura \& Sunyaev }{1973}]{SS73}
Shakura, N.~I., \& Sunyaev, R.~A.\ 1973, \aap, 24, 337

\bibitem[\protect\citeauthoryear{Steffen et al.}{2010}]{steffen10}
Steffen, J.~H., Batalha, N.~M., Borucki, W.~J., et al. 2010, \apj, 725, 1226


\bibitem[\protect\citeauthoryear{Tanaka et al.}{2002}]{Tan02}
Tanaka, H., Takeuchi, T., \& Ward, W.~R.\ 2002, \apj, 565, 1257

\bibitem[\protect\citeauthoryear{Terquem \& Papaloizou}{2007}]{Ter07}
Terquem, C., \& Papaloizou, J.~C.~B.\ 2007, \apj, 654, 1110

\bibitem[\protect\citeauthoryear{Vorobyov \& Basu}{2009}]{Vor09}
Vorobyov, E.~I., \& Basu, S.\ 2009, \apj, 703, 922

\bibitem[\protect\citeauthoryear{Ward}{1997}]{Ward97}
Ward, W.~R.\ 1997, Icarus, 126, 261

\bibitem[\protect\citeauthoryear{Wang \& Zhou}{2011}]{WZ11}
Wang, S., \& Zhou, J.~L.\ 2011, \apj, 727, 108

\bibitem[\protect\citeauthoryear{Zhang et al.}{2010}]{Zhang10}
Zhang, N., Ji, J., \& Sun, Z.\ 2010, \mnras, 405, 2016


\bibitem[\protect\citeauthoryear{Zhou et al.}{2005}]{Zhou05}
Zhou, J.~L., Aarseth, S.~J., Lin, D.~N.~C., \& Nagasawa, M.\ 2005,
\apjl, 631, L85

\bibitem[\protect\citeauthoryear{Zhou \& Lin}{2008}]{ZL08}
Zhou, J.~L., \& Lin, D.~N.~C.\ 2008, IAU Symposium, 249, 285

\end{thebibliography}
\end{document}